\documentclass[12pt]{article}
\usepackage{amsmath, amsthm}
\usepackage{url}
\usepackage{graphicx,psfrag,epsf}
\usepackage{enumerate}
\usepackage{natbib}
\usepackage{verbatim}
\usepackage{jmlr2e}
\setcitestyle{numbers,open={[},close={]}}

\usepackage[symbol]{footmisc}
\usepackage{epstopdf,gensymb}
\usepackage{graphics}
\usepackage{graphicx,color}

\usepackage{amsfonts, animate,subfigure}
\usepackage{amssymb}
\usepackage[margin=1in]{geometry}
\usepackage{setspace}
\usepackage{indentfirst}
\usepackage{float}
\usepackage{amsfonts}
\usepackage{bm,pbox}
\usepackage{multirow} 
\usepackage{color}
\usepackage[margin=1in]{geometry}
\usepackage{framed}
\usepackage{amssymb,amsthm,amsfonts,amsbsy,latexsym,amsmath}
\usepackage{enumerate}

\newtheorem{theorem}{Theorem}

\def\qed{ \hfill $\blacksquare$}  
\begin{document}
	\def\spacingset#1{\renewcommand{\baselinestretch}%
	{#1}\small\normalsize} \spacingset{1}
	\title{\bf The Power of A/B Testing under Interference\thanks{~~ D.T.U  gratefully acknowledges the National Science Foundation grant NSF DMS-1312361 for providing partial support for his work on this manuscript. 
}\hspace{.2cm}}
	  \author{\name James D. Wilson \email jdwilson4@usfca.edu \\
	  \addr Department of Mathematics and Statistics\\
	University of San Francisco\\
	San Francisco, CA 94117 
	\AND
	\name David T. Uminsky  \email duminsky@usfca.edu \\
	  \addr Department of Mathematics and Statistics\\
		University of San Francisco\\
		San Francisco, CA 94117 
	}
		
\maketitle

\begin{abstract}
	{\footnotesize In this paper, we address the fundamental statistical question: how can you assess the power of an A/B test when the units in the study are exposed to interference? This question is germane to many scientific and industrial practitioners that rely on A/B testing in environments where control over interference is limited. We begin by proving that interference has a measurable effect on its sensitivity, or power. We quantify the power of an A/B test of equality of means as a function of the number of exposed individuals under any interference mechanism. We further derive a central limit theorem for the number of exposed individuals under a simple Bernoulli switching interference mechanism. Based on these results, we develop a strategy to estimate the power of an A/B test when actors experience interference according to an observed network model. We demonstrate how to leverage this theory to estimate the power of an A/B test on units sharing any network relationship, and highlight the utility of our method on two applications - a Facebook friendship network as well as a large Twitter follower network. These results yield, for the first time, the capacity to understand how to design an A/B test to detect, with a specified confidence, a fixed measurable treatment effect when the A/B test is conducted under interference driven by networks.}
\end{abstract}

\noindent%
\newpage
\spacingset{1} 



A/B testing is a fundamental inferential technique used to assess the difference between two classes of individuals that are exposed  to different treatments. A/B testing has been extensively used for case studies in infectious disease and clinical testing \cite{rubin1980randomization}, as well more modern applications such as recommendation systems \cite{Davidson:2010}, gaming \cite{Andersen:2011}, personalized search \cite{Hannak:2013}, and e-commerce \cite{Kohavi:2007}. These complex applications have stimulated demands for massive online experiments that require both high volume and velocity of data. As a result, new theoretical and computational techniques are increasingly an important areas of research \cite{Kohavi:2013}.

Formally, an A/B testing experiment is an experimental design used to quantify the difference in treatments on two independent samples of actors. Members from opposing samples undergo different treatments, and the \emph{treatment effect} is assessed by testing the equality of some continuous or discrete measurement on the two samples. To be more precise, consider a sample of $n$ actors. First, $n_A$ members are assigned (randomly or deterministically) to class $A$ and the remaining $n_B = n - n_A$ are assigned to class $B$. Each actor $j \in \{1, \ldots, n\}$ is given a class label $c_j \in \{A, B\}$ according to the class that he or she is assigned. Subsequently, an experiment is performed wherein members of class $A$ are given different treatments or exposures than members of class $B$, and a measurement $x_j$ is recorded for actor $j$.


Each measurement $x_j$ is viewed as an independent draw from a probability density or mass function $f_k(\cdot)$ with finite mean $\mu_k$, where $k \in \{A,B\}$ is the class label of actor $j$. Commonly, one assesses the difference between the two treatments on the two samples by estimating the average treatment effect (ATE), $\delta = \mu_A - \mu_B$ using the data $\{x_j: j = 1, \ldots, n\}$. One deems whether or not the average treatment effect is statistically significant by testing the null hypothesis

\begin{equation}\label{eq:hypothesis_test}
	H_0: \delta = 0. 
\end{equation}

The validity of (\ref{eq:hypothesis_test}) depends upon the stable unit treatment value assumption (SUTVA) \cite{rubin1980randomization}, under which, among other things, there is no \emph{interference} among the outcomes of actors in the study. Interference occurs when the treatment of an individual is altered due to interactions of that actor with other individuals in the study \cite{hudgens2012toward, tchetgen2012causal}. However, when actors interact with one another in a social network, SUTVA is often violated \cite{gui2015network, Xu:2015}. If interference is not appropriately accounted for, (\ref{eq:hypothesis_test}) is no longer valid since one cannot rely on the measurements used to assess treatment effects of each class.

An important component of understanding (\ref{eq:hypothesis_test}) under interference involves the development and analysis of unbiased estimators for the ATE $\delta$ and other related treatment effect quantities used for causal inference. The seminal works \cite{hudgens2012toward, tchetgen2012causal} provided unbiased causal estimands for $\delta$ under a general stratified design. The authors of \cite{aronow2012estimating} generalized these results using a Horvitz-Thompson estimation strategy based on the knowledge of which participants were exposed to interference. The recent work of \cite{basse2017limitations} analyzed the limitations of unbiased causal estimators and showed that the variance of such estimators do not, in general, shrink to zero when interference is introduced in an A/B test. This work showcased the fact that common strategies for causal inference cannot be adapted to the situation under which interference occurs. 

Despite the recent work done in estimation of causal effects, no work has been done on understanding the effects of interference on testing. In light of this, we  analyze, for the first time, the effects of interference on the A/B test in (\ref{eq:hypothesis_test}). We investigate two important related questions:

\begin{enumerate}
	\item To what extent does known interference affect the sensitivity and specificity of an A/B test? 
	\vskip .5pc
    \item How can one estimate the power of an A/B test when the actors undergo interference?
\end{enumerate}

We address these two questions by first showing that interference has a measurable effect on the an A/B test's sensitivity. We then characterize the power of an A/B test of equality of means as a function of the number of exposed individuals under any interference mechanism. We next derive a central limit theorem for the number of exposed individuals under a simple Bernoulli switching interference mechanism. Based on these results, we develop a strategy to estimate the power of an A/B test when actors experience interference according to an observed network model, and demonstrate how to leverage this theory to estimate the power of an A/B test on units sharing any network relationship. Our method is lastly applied to two applications - a Facebook friendship network and a large Twitter follower network.


\subsection*{Motivating Example and previous approaches}
A/B testing has widely become the standard controlled experimentation framework for network driven companies like Facebook, LinkedIn, Twitter, Google, and Yahoo despite the fact that the key SUTVA assumption is in clear violation. At these companies, behavior of a user has a likely and measurable impact on the users connected in his or her social or professional neighborhood on the network.

To see this, imagine the setting in which Twitter has decided to test a new cat feature to understand the effect of this feature on a user's engagement time on Twitter. To conduct a standard A/B test, Twitter randomly assigns its users into one of two classes, either A or B. Treatment group A is given their usual Twitter feed, and group B is given the ``cat feature,'' which introduces cat videos to the user's feed at a random but regular rate.  User engagement is then measured for each group. For each class, the sample mean of active engagement time is calculated. Using these sample means, the company can then statistically quantify the effect of these strategies one tests the equality of the population active time on the app for each strategy - $\mu_k, k \in \{A,B\}$ - using (\ref{eq:hypothesis_test}). 


Interference intrinsically occurs in this example. User engagement of a treatment B user includes reposting or "tweeting" content, which may include the new cat videos from the ``cat feature.'' Subsequently, followers of this user may be exposed to the cat videos, including users who are in class A.

Not only is SUTVA violated in this example, but an active user given the no cat treatment (A) but who's social network has active members getting the  cat treatment  can, in affect ``switch'' or change the treatment from $A$ to $B$ due to his or her friends' posts.


This overly simple example provides an illustration of one of the core network interference effects on A/B Testing. The obvious approach of ``turning off" a user's social network during an experiment fails for many applications who's standard function and interaction metrics are highly dependent on that aspect of the application. In effect, turning off social media functionality will cause different user behavior even before an A/B test can be conducted. 

This is a significant problem and has been recently dubbed the {\em Network A/B Test} problem \cite{gui2015network}. There has been a lot of approaches taken to understand, model, correct and redesign A/B tests in this setting. In \cite{Sobel:2006,hudgens2012toward,aronow2012estimating}, the researchers focus on the effect of interference on ATE as well as  develop out a framework to understand the effect of interference  on causal inference conclusions for an A/B test. In \cite{Eckles:2016}, new theory were developed to decrease bias in estimators that result from network interference. Authors in \cite{backstrom2011network,Katzir:2012} use a network sampling approach called network bucketing, to decrease the bias induced by network interference. 


\section*{Results}
\subsection*{A/B Testing under Interference}
We first investigate the effects of interference described above on the sensitivity and specificity of inference made on (\ref{eq:hypothesis_test}). We consider the following mechanism of interference:

\begin{framed}
\noindent {\bf Interference Mechanism}
\begin{itemize} 
\item Labels $\mathbf{c} \in \{A, B\}^{n}$ are chosen randomly or deterministically, and actors are assigned to treatments $A$ or $B$ according to his or her class label.
\item A subset of actors are exposed to interference and their class labels are altered. The new \emph{unknown} class labels are ${\mathbf{d}} \in \{A, B\}^{n}$.
\item Actors receive treatments according to class labels $\mathbf{d}$. 
\end{itemize}  
\end{framed}

Measurements are made on each actor and represented by $\{x_j: j = 1, \ldots, n\}$. The sample mean for each class $\overline{x}_A = \text{Ave}(\{x_j: c_j = A\})$ and $\overline{x}_B = \text{Ave}(\{x_j: c_j = B\})$ and a decision on whether or not to reject H$_0$ is made according to the distribution $f_k(\cdot)$ and the test statistic

\begin{equation}\label{eq:test_stat} T_{\mathbf{c}} = \dfrac{\overline{x}_A - \overline{x}_B}{\sqrt{(\sigma_A^2/n_A + \sigma_B^2/n_B)}},\end{equation}

\noindent where $\sigma^2_k$ and $n_k$ is the population variance and size of class $k$ under labels $\mathbf{c}$, respectively. For the alternative hypothesis that $\delta > 0$, one decides to reject $H_0$ when $T_{\mathbf{c}} > m$ for some $m$ chosen to control the sensitivity and specificity of the test, where sensitivity is the long run probability of correctly rejecting the null hypothesis, and specificity is the probability that the null hypothesis is correctly supported. When SUTVA holds, one can directly calculate these quantities under the originally specified and known labels $\mathbf{c}$. However, in the case that interference has occurred according to the interference mechanism above, we instead must account for the true (but unknown) labels $\mathbf{d}$ while evaluating null hypothesis (\ref{eq:hypothesis_test}) under $\mathbf{c}$. In particular, we assess the following quantities:

\begin{equation}\label{typeI}
\gamma_{\mathbf{c}|\mathbf{d}} = \text{Pr}(\text{Reject } \text{H}_{0} \text{~under~} \mathbf{c} \mid \text{H}_0, \mathbf{d} \text{~true})
\end{equation}

\begin{equation}\label{typeII}
\beta_{\mathbf{c}|\mathbf{d}} = \text{Pr}(\text{Reject } \text{H}_{0} \text{~under~} \mathbf{c} \mid \text{H}_0 \text{~false}, \mathbf{d} \text{~true}).
\end{equation}


Above $\gamma_{\mathbf{c}|\mathbf{d}}$ is the probability of incorrectly rejecting the null hypothesis under label misspecification. The quantity $1 - \gamma_{\mathbf{c} | \mathbf{d}}$ measures the specificity of the test under misspecification. The quantity $\beta_{\mathbf{c}|\mathbf{d}}$ is the \emph{power} of the test and is a measure of the sensitivity of the test. Since we do not know the true labels ${\bf d}$, we evaluate (\ref{typeI}) and (\ref{typeII}) as a function of how many assignments remain the same as our original specification in $\mathbf{c}$. Accordingly, define

$$n_S = \sum_{j = 1}^{n} \mathbb{I}(c_j = d_j)$$

\noindent as the number of actors that keep the same label after interference, and $n_D = n - n_S$ as the number of actors who changed class assignments due to interference. We consider two cases that are commonly used in the context of A/B testing, including the case that $f_k(\cdot)$ is a normal distribution $N(\mu_k, \sigma^2)$ and the case that $f_k(\cdot)$ is the Bernoulli distribution $Bernoulli(\mu_k)$. 

In each of the cases below, we let $Z_\alpha$ denote the critical value of the Normal distribution that solves $1 - \Phi(z) = \alpha$, where $\Phi(\cdot)$ is the cumulative distribution function of a N(0,1) random variable and $\alpha \in (0,1)$.

\subsubsection*{Case I: Under Normality}
We first evaluate the simplest case of test (\ref{eq:hypothesis_test}), under which the observed measurements $\{x_j: j = 1, \ldots, n\}$ are random draws from a mixture of normal distributions with known variance $\sigma^2 > 0$. We reject H$_0$ if $T_{\mathbf{c}} > Z_\alpha$, where $T_\mathbf{c}$ is the test statistic in (\ref{eq:test_stat}). Our first result describes the effect of class misspecification in this scenario.

\begin{theorem}\label{thm:Normal_case}
Suppose that $\{x_j: j = 1, \ldots, n\}$ are random draws from a N($\mu_j$, $\sigma^2$) distribution, where 
$$\mu_j = \mu_A ~ \mathbb{I}(d_j = A) + \mu_B ~ \mathbb{I}(d_j = B).$$ 
Let $\mathbf{c} \neq \mathbf{d}$ be a class assignment that has $n_S$ assignments in common with $\mathbf{d}$ and $n_D = n - n_S$ assignments that differ. Suppose that $n_A = n_B = n/2$. Then the following hold.
\begin{itemize}
	\item(){\rm (a)} Under test (\ref{eq:hypothesis_test}),
	\begin{align}\label{thm1_type1}
    \gamma_{\mathbf{c} \mid \mathbf{d}} = \gamma_{\mathbf{d} \mid \mathbf{d}}
    \end{align}
	\vskip .5pc
	\item(){\rm (b)} Suppose that the alternative hypothesis to $H_0$, given by $H_1: \mu_A > \mu_B$, is true and $\mu_A - \mu_B = \delta > 0$. Then for all finite $n$,
	\begin{align}\label{thm1_type2} \beta_{\mathbf{c} \mid \mathbf{d}} &= 1 - \Phi\left(Z_\alpha - \dfrac{(n_S - n_D) \delta}{2\sigma\sqrt{n}}\right) 
	\end{align}
\end{itemize}
\end{theorem}

Theorem 1 describes the consequences of running an A/B test without accounting for interference under the test of difference of means. Equation (\ref{thm1_type1}) reveals that the specificity of an A/B test is not affected by interference. Equation (\ref{thm1_type2}) provides an exact expression for the calculation of the power of the test under misspecification. This probability is a function of the number of misspecified labels. Indeed, the power of the test is a monotonically decreasing function in the number of misspecified assignments. Thus as interference is increased, the power of the test decreases.

\subsubsection*{Case II: Difference of Proportions}

We now evaluate the case where the measurement $x_j$ is an indicator variable that designates a success or failure, such as whether or not an actor clicks an ad. In this case, the observations $\{x_j: j = 1, \ldots, n\}$ are modeled as random draws from a mixture of Bernoulli distributions with the success proportion depending on the class of the actor. We reject H$_0$ if $T_{\mathbf{c}} > Z_\alpha$, where $T_\mathbf{c}$ is the test statistic in (\ref{eq:test_stat}). We will write $a_n \approx b_n$ to denote that $a_n / b_n \rightarrow 1$ as $n \rightarrow \infty$. 

\begin{theorem}\label{thm:Bernoulli_case}
	Suppose that $\{x_j: j = 1, \ldots, n\}$ are random draws from a Bernoulli($\mu_j$) distribution, where 
	$$\mu_j = \mu_A ~ \mathbb{I}(d_j = A) + \mu_B ~ \mathbb{I}(d_j = B).$$ 
	Let $\mathbf{c} \neq \mathbf{d}$ be a class assignment that has $n_S$ assignments in common with $\mathbf{d}$ and $n_D = n - n_S$ assignments that differ. Suppose that $n_A = n_B = n/2$. Suppose that the alternative hypothesis to $H_0$, given by $H_1: \mu_A > \mu_B$, is true and $\mu_A - \mu_B = \delta > 0$. Then as $n \rightarrow \infty$,
		\begin{equation}\label{thm2_type2}
      \beta_{\mathbf{c} | \mathbf{d}} \approx 1 - \Phi\left(\sqrt{\dfrac{n}{2}} \dfrac{Z_\alpha}{\sigma_{A,B}} - \dfrac{(n_S - n_D) \delta}{2\sqrt{\widetilde{n}_A\sigma_A^2 + \widetilde{n}_B \sigma_B^2}} \right),
		\end{equation}
	\noindent where $\sigma_{A,B}^2 = \dfrac{\sigma_A^2 + \sigma_B^2}{\widetilde{n}_A \sigma_A^2 + \widetilde{n}_B\sigma_{B}^2}$, $\sigma_k^2 = \mu_k(1-\mu_k)$, and $\widetilde{n}_k = \sum_j \mathbb{I}(d_j = k)$.
\end{theorem}
Equation (\ref{thm2_type2}) shows that the approximate power function is again a function of $n_S - n_D$. The power of the test monotonically decreases as the number of misspecified labels $n_D$, and hence the effect of interference, increases. For fixed $\delta > 0$ and $n_S - n_D = o(\sqrt{n})$, the power converges to 0 as $n$ tends to $\infty$. On the other hand, when $n_S - n_D = \Omega(\sqrt{n})$, the power converges to 1, suggesting that for large $n$ the power of the A/B test performs similarly as A/B tests without interference. 

This result provides intuition for the rate at which the power of an A/B test is changed under class misspecification. When the number of correctly specified labels, $n_S$, is sufficiently large the power, and hence the sensitivity of the test is not strongly affected by interference.
\section*{Estimating Power under Interference}
\begin{figure}
	\centering
    \caption{\label{fig:power} Expected power of A/B testing under Bernoulli switching. The top plots show expected power as a function of the sample size $n$ and the bottom plots illustrate the expected power as a function of the effect size $\delta$.}
    \includegraphics[trim = 0cm 0cm 0cm 0cm, clip = TRUE, width = \textwidth]{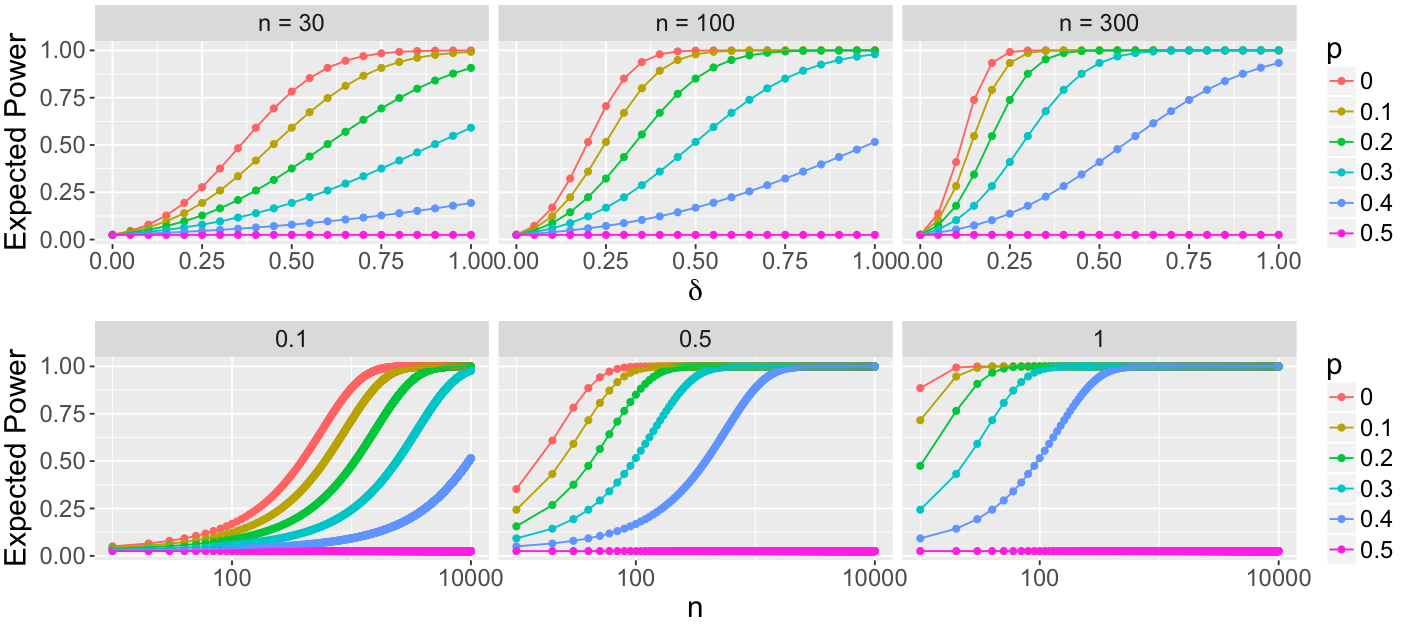}
\end{figure}

Results (\ref{thm1_type2}) and (\ref{thm2_type2}) show that under Normality and the case of proportions, the power of an A/B test is a function of the number of misclassified labels,  $n_S - n_D$. Thus, getting a handle on $n_S - n_D$ enables the determination of the approximate power of an A/B test on the population. That is, if $n_S - n_D$ can be exactly determined, one can use (\ref{thm1_type2}) or (\ref{thm2_type2})  to quantify the power of an A/B test. Often, one cannot immediately calculate $n_S - n_D$ because the knowledge of who received the incorrect treatment is unknown. In cases like this, one must consider the mechanism for \emph{who} will receive the wrong class assignment or treatment. For this, one can obtain the probability of each actor succumbing to interference. This problem is akin to approximating the number of infected people in an epidemic spreading process \cite{kempe2003maximizing,PhysRevLett.92.218701,hodas2014simple}. 

We analyze a simple stochastic mechanism for interference, which we will refer to as Bernoulli switching. Under this mechanism, each actor $i$ in the system independently receives the incorrect treatment with probability $p_i$. One can readily approximate the value of $n_S - n_D$ under Bernoulli switching using the following central limit theorem.

\begin{theorem}\label{thm:CLT}
Suppose that the $n$ actors in the system receive treatments under which interference occurs according to the Bernoulli switching mechanism. That is, each actor receives the opposite treatment than what he or she was assigned independently with probability $p_i$. Let $s^2_n = \sum_{i = 1}^n p_i(1-p_i)$. As $n \rightarrow \infty$, if $s_n \rightarrow \infty$ then

\begin{equation}\label{eq:CLT}
\dfrac{n_S - n_D - \mu_{\bf p}}{2s_n}  \stackrel{D}{\rightarrow} N(0, 1),
\end{equation}
\noindent where $\mu_{\bf p} = n - 2\sum_{i = 1}^n p_i$.
\end{theorem}

Theorem \ref{thm:CLT} suggests a straightforward approach to estimating the power of an A/B test under interference for an observed system of $n$ actors and an underlying network describing the actors relationships.
\newpage
\begin{framed}
\noindent {\bf Power Estimation Procedure}\\
\noindent \emph{Given}: a collection of actors $(n) = \{1, \ldots, n\}$, their interference network $G = ((n), E)$, and class labels ${\bf c}$
\begin{enumerate}
\item Calculate the probability that each actor receives the incorrect (and opposite) treatment. For example, in the case of social interference, one may estimate $p_i$ according to the class labels of actor $i$'s neighbors in $G$.
\item Estimate $n_S - n_D$ as well as $\widetilde{n}_A$ and $\widetilde{n}_B$. 
\item Estimate the power of the test using (\ref{thm1_type2}) or (\ref{thm2_type2}). 
\end{enumerate}
\end{framed}

This approach provides a direct manner to estimate the power of an A/B test on actors who may undergo interference. This procedure provides an assessment of the reliability of the results of an A/B test, and in so doing, acts as a yard stick as to whether or not intervention of some kind is needed in the devising of the experiment. In the case of the Bernoulli switching mechanism, one can readily estimate $n_S - n_D$ as well as $\widetilde{n}_A$ and $\widetilde{n}_B$ using the central limit theorem result from (\ref{CLT_1}). As an example of how one can use these results in practice, we consider the expected power of an A/B test with equally sized classes, where actors undergo Bernoulli switching with probability $p$. We illustrate these results in Figure \ref{fig:power}, and show how the expected power changes as a function of $n$ and $\delta$. These plots provide an approximation of what a practitioner can expect in a network with varying degrees of interference. In the next section, we further explore the utility of our proposed method through its application to two social networks from Facebook and Twitter.

\section*{Applications}

\begin{figure}[H]
	\centering
    \caption{\label{fig:degree_dists} The degree distribution of the Facebook and Twitter networks considered in this application study. Both networks demonstrate behavior of a scale-free social network.}
    \includegraphics[trim = 0cm 0cm 0cm 0cm, clip = TRUE, width = 0.6\textwidth]{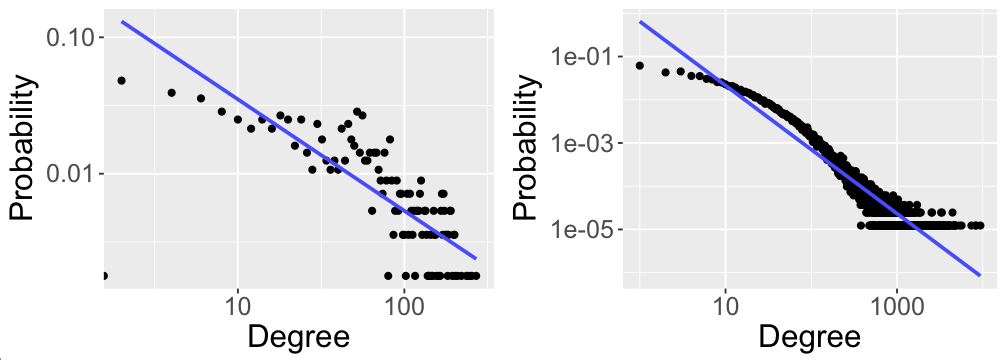}
\end{figure}

We now apply our power estimation procedure on two social networks: the Facebook friendship network from \cite{wilson2014testing} and the Twitter social circles network from \cite{leskovec2012learning}. Members of each of these networks commonly undergo interference due to the spread of information and media among social groups. The Facebook network contains 561 nodes representing Facebook members and 16750 undirected edges, which represent friendships among the actors. The Twitter network contains 81306 nodes and 2420766 directed edges, where edges point to actors that are followed and point from followers. The degree distributions of each of these networks are plotted in Figure \ref{fig:degree_dists}. As observed in Figure \ref{fig:degree_dists}, the degree distributions of each of these networks demonstrate typical behavior of scale-free social networks, as seen by the linear relationship between the log degree and log probabilities.  

We model contagion on each of these networks in the following way. First $\left \lceil{n~p.A}\right \rceil$ of the actors are randomly assigned label $A$ and the remaining nodes are assigned label $B$. These labels are stored in the vector ${\bf c} = \{c_1, \ldots, c_n\}$. Given ${\bf c}$, nodes simultaneously look at their neighbors. A node changes its class label with probability equal to the proportion of its neighbors that have the opposite label. Let $\mathcal{N}(i)$ denote the neighborhood of node $i$. Then actor $i$ switches classes with probability
$${p}_i = \dfrac{|\{u \in \mathcal{N}(i): c_u \neq c_i \}|}{|\mathcal{N}(i)|}$$

\begin{figure} [H]
	\centering
    \caption{\label{fig:exp_power} The expected power plots of A/B testing for the Facebook and Twitter networks as a function of the proportion of individuals in class A ($p.A$) and the effect size $\delta$.}
      \includegraphics[trim = 0cm 0cm 0cm 0cm, clip = TRUE, width = 0.9\textwidth]{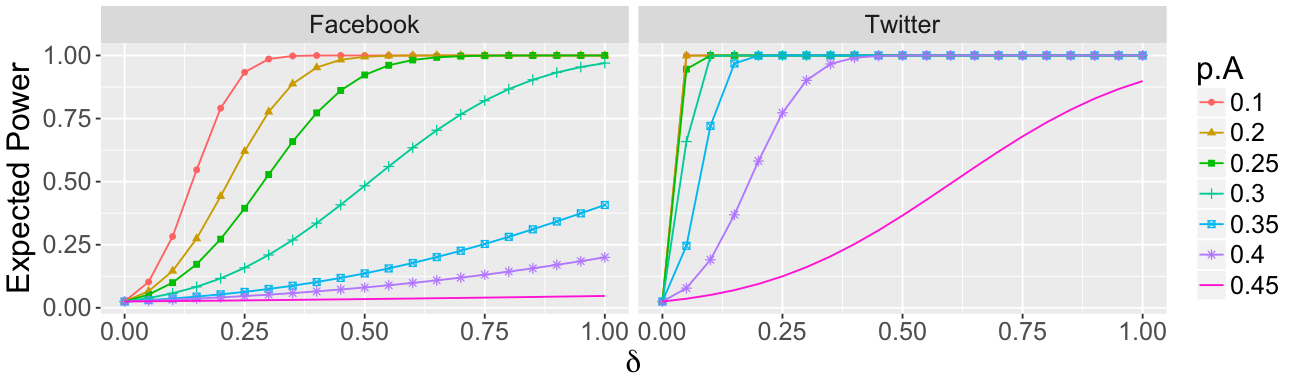}
\end{figure}

This model of interference is closely related to the contagion spread according to the Watt's Threshold model \citep{watts2002simple}. This model assumes that a node's neighbors are the most influential on the treatment he or she receives. Using $\mathbf{p}  = \{p_1, \ldots, p_n\}$, we estimate $n_S - n_D$ by its mean $\mu_{\mathbf{p}}$ from result (\ref{eq:CLT}), and estimate the power of an A/B test under normality as specified in Theorem 1. We set $\mu_A = 0$, $\mu_B = 1$, and $\sigma = 1$ and calculate the expected power over a grid of values of $\delta$ and $p.A$. The results for each of these networks are shown in Figure \ref{fig:exp_power}. 

Figure \ref{fig:exp_power} reveals that the proportion of nodes in each class has a dramatic effect on the power of an A/B test in these networks. Across all values of $\delta$, power is monotonically decreases as $p.A$ increases to 0.5. This result is due to the neighborhood interference mechanism that we apply. Note that for a randomly selected node, the likelihood of the node changing its class is proportional to $p.A (1-p.A)$, which is maximized when $p.A = 0.50$. We also observe the effects of network size and effect size $\delta$ on the power of the A/B test with these two applications. The Twitter network has much higher power across all values of $p.A$ and $\delta$ than the Facebook network, which has less than 1/100th of the nodes. These plots provide a diagnostic for which practitioners can assess the strength of an A/B test on when interference occurs through a network. 


\section*{Discussion}
The theoretical results and applications presented here provide an initial understanding of the effects of interference on the power of A/B tests. These new, `off the shelf' techniques can be applied to existing testing frameworks in both research and industry environments. The results of this paper open up many more questions and opportunities for further research, which we discuss here.

While Theorems 1 and 2 cover differences of means and proportions under normal and Bernoulli measurements, analyzing measurements under other parametric distributions is an important area of future work. Although Theorems 1 and 2 quantify the power of the A/B test under any interference mechanism based on the number of exposed individuals $n_D$, we analyze the distribution of $n_D$ under the Bernoulli switching interference mechanism which does not directly account for interactions among actors like that occurring in a social network. The results in Theorem 3 can and should be generalized to such network interference settings. One can analyze interference mechanisms like the Watts threshold model \citep{watts2002simple} on differing families of random graph models like the configuration model or the stochastic block model. In situations where the dependence among actors is weak - for example in sparse random graphs where the total degree of the network grows much slower than the size of the network, we expect a central limit theorem like that of Theorem 3 to hold. 


Finally, incorporating the existence of network interference that causes a new intermediate treatment group C or studying the more general experimentation framework of more than two treatment groups remains an open and important area of theoretical development. Along these lines, one can use the results presented here to explore a continuum of contagion rather than a complete switch like that considered in this paper. For industry researchers who have access to the full network dynamics, it may be possible to infer both the appropriate mechanisms and parameters of such models from their user behavior. Users can then leverage more accurate contagion models (based on their own network behavior) to better inform the design of A/B experiments for specified power requirements.



\section*{Proofs}
\subsection*{Proof of Theorem 1}
To prove statements (a) and (b), we assess the distribution of $T_{\bf c} = \frac{\overline{x}_B - \overline{x}_A}{\sqrt{4\sigma^2 / n}}$ under the null and alternative hypothesis, respectively. Here, $\overline{x}_k = \text{Ave}(\{x_j: c_j = k\})$ and the labels are misspecified due to interference. Let $n_{k\ell} = \#\{\text{$j$: $c_j = k$ and $d_j = \ell$}\}$. Then we first assess the distribution of $\overline{x}_B - \overline{x}_A$ under label misspecification. Note that we can first decompose this difference in the following way

\begin{align}\label{eq:decompose}
	\dfrac{n}{2}(\overline{x}_B - \overline{x}_A) & = \sum_{c_j = d_j = B} x_j + \sum_{c_j = B, d_j = A} x_j - \sum_{c_j = d_j = A} x_j - \sum_{c_j = A, d_j = B} x_j
\end{align}

\noindent Equation (\ref{eq:decompose}) yields
\begin{align*}
\text{Var}\left(\dfrac{n}{2}(\overline{x}_B - \overline{x}_A)\right) &= n_{BB} \sigma^2 + n_{BA}\sigma^2 + n_{AA}\sigma^2 + n_{AB}\sigma^2 = n\sigma^2.
\end{align*}

\noindent Furthermore,
\begin{align*}
\mathbb{E}\left(\dfrac{n}{2}(\overline{x}_B - \overline{x}_A)\right) & = n_{BB}\mu_B + n_{BA}\mu_A - n_{AA}\mu_A - n_{AB}\mu_B\\
& = n_{BB}\mu_B + n_{BA}(\mu_B - \delta) \\
& - n_{AA}\mu_A - n_{AB}(\delta + \mu_A)\\
& = \dfrac{1}{2}(n_S - n_D) \delta.
\end{align*}

\noindent Since $x_j$ are each normally distributed, it follows that $\dfrac{n}{2}(\overline{x}_B - \overline{x}_A)$ is normally distributed with mean and variance given above. It follows that under $H_1$, $T_{\bf c}$ is distributed as $N\left(\dfrac{(n_S - n_D)\delta}{2\sigma\sqrt{n}}, 1\right)$. When $H_0$ is true and $\delta = 0$, the distribution of $T_{\bf c}$ does not depend on the misspecified labels. Statement (a) of the theorem follows. Now, let $Z$ be a standard normal random variable. When $H_1$ is true, we have that the power under misspecification is 
\begin{align*}
\beta_{\bf c \mid \bf d} &= \text{Pr}(T_{\bf c} > Z_\alpha \mid H_1 ~\text{true})\\
& = \text{Pr}\left(Z > Z_\alpha - \dfrac{(n_S - n_D) \delta}{2\sigma \sqrt{n}}\right)\\
& = 1 - \Phi\left(Z_\alpha - \dfrac{(n_S - n_D) \delta}{2\sigma \sqrt{n}} \right).
\end{align*}
This completes the proof. \qed

\subsection*{Proof of Theorem 2}
We proceed with this proof in the same manner as the proof of Theorem 1 above by first determining the mean and variance of $\dfrac{n}{2}(\overline{x}_B - \overline{x}_A)$. The decomposition in (\ref{eq:decompose}) leads to 
\begin{align*}
\mathbb{E}\left(\dfrac{n}{2}(\overline{x}_B - \overline{x}_A) \right) = \dfrac{1}{2}(n_S - n_D) \delta, \end{align*}
\noindent and
\begin{align*}
\text{Var}\left(\dfrac{n}{2}(\overline{x}_B - \overline{x}_A)\right) = \widetilde{n}_A \sigma^2_A + \widetilde{n}_B \sigma^2_B,
\end{align*}
\noindent where $\sigma^2_k = \mu_k(1 - \mu_k)$. The test statistic from (\ref{eq:test_stat}) reduces to $T_{\mathbf{c}} = \frac{\overline{x}_A - \overline{x}_B}{\sqrt{2/n(\sigma_A^2 + \sigma_B^2)}}$. Under $H_1$, we have that the approximate distribution of $T_{\mathbf{c}}$ is $N\left(\dfrac{(n_S - n_D) \delta}{\sqrt{2n(\sigma_A^2 + \sigma_B^2)}}, \dfrac{2}{n} \sigma^2_{A,B}\right)$. It follows that the power under label misspecification is
\begin{align*}
\beta_{\bf c \mid \bf d} &= \text{Pr}(T_{\bf c} > Z_\alpha \mid H_1 ~\text{true})\\
& \approx 1 - \Phi\left(\sqrt{\dfrac{n}{2}} \dfrac{Z_\alpha}{\sigma_{A,B}} - \dfrac{(n_S - n_D) \delta}{2\sqrt{\widetilde{n}_A\sigma_A^2 + \widetilde{n}_B \sigma_B^2}} \right).
\end{align*}
This completes the proof. \qed

\subsection*{Proof of Theorem 3}

Let ${\bf X} = (n_S, n_D)^T$. Under Bernoulli switching, linearity of expectation gives $\mathbb{E}(n_D) = \sum_{i = 1}^n p_i$ and $\text{Var}(n_D) = \sum_{i = 1}^n p_i(1-p_i) = s_n^2$. Analogously, $\mathbb{E}(n_S) = n - \sum_{i = 1}^n p_i$, $\text{Var}(n_S) = \text{Var}(n_D)$, and $\text{Cov}(n_S, n_D) = -\sum_{i = 1}^n p_i(1-p_i) = -s_n^2$.

Since label switching occurs independently across nodes, if $s_n \rightarrow \infty$ as $n \rightarrow \infty$, then Lindeberg's central limit theorem condition \cite{billingsley2013convergence} is met and the following multivariate central limit theorem holds:

\begin{equation}\label{CLT_1}
\Sigma^{-1/2}\left({\bf X} - \mathbb{E}({\bf X})\right) \stackrel{D}{\rightarrow} N({\bf 0}, \mathbb{I}_2),
\end{equation}

\noindent where $\mathbb{I}_2$ is the two dimensional identity matrix and

$$\Sigma = \begin{pmatrix} s_n^2 & -s_n^2 \\ -s_n^2 & s_n^2 \end{pmatrix}, \qquad \mathbb{E}({\bf X}) = \begin{pmatrix} n - \sum_{i=1}^np_i \\ \sum_{i = 1}^n p_i \end{pmatrix}.$$

Let ${\bf r} = (1, -1)^T$. Note that $n_S - n_D = {\bf r}^T {\bf X}$. Applying the multivariate delta method to (\ref{CLT_1}) to the distribution of ${\bf r}^T{\bf X}$ yields the desired result. \qed


\end{document}